\documentstyle[sprocl,psfig]{article}

\def\Journal#1#2#3#4{{#1} {\bf #2}, #3 (#4)}

\def\PLB{{\em Phys. Lett.}  B}

\def\PRD{{\em Phys. Rev.} D}

\def\ZPC{{\em Z. Phys.} C}

\def\be{\begin{equation}}
\def\ee{\end{equation}}
\def\bea{\begin{eqnarray}}
\def\eea{\end{eqnarray}}

\begin{document}

\title{HADRONIC FLUCTUATIONS AND QUARK-ANTIQUARK ASYMMETRY IN THE NUCLEON
\footnote{Presented in {\it Quark Matter'97}, Tsukuba, Japan, and in 
{\it Hadron Physics'98}, Florianopolis, Brazil.}}
\author{H.R. CHRISTIANSEN and J. MAGNIN}

\address{Centro Brasileiro de Pesquisas F\'{\i}sicas, CBPF - DCP \\ 
Rua Dr. Xavier Sigaud 150, 22290-180, 
Rio de Janeiro, Brazil\\E-mail: hugo@cat.cbpf.br, jmagnin@lafex.cbpf.br} 

\maketitle\abstracts{ We derive the nucleon non-perturbative sea-quark 
distributions coming from a composite model involving quarks and hadronic 
degrees of freedom. The model predicts a definite structured $q-\bar q$ 
asymmetry in the nucleon sea.}

\section{Introduction}
The quark-antiquark sea of a hadron is mainly generated in a perturbative 
way as $Q^2$ rises up. Nevertheless, it is widely believed that some 
fraction of the sea quarks may be associated with non-perturbative 
processes. One of the most important issues of such a prospect is the 
possibility of having non-symmetric quark-antiquark distributions which 
could, in principle, be measurable. This is a topical 
question related to current experiments concerning either
light or heavy flavors.
 For instance, regarding the strange sea of the nucleon, the present 
status of the experimental data does not exclude the possibility of 
asymmetric $s$ and $\bar{s}$ distributions. The CCFR Collaboration 
\cite{ccfr} has analyzed the strange quark distribution in protons 
allowing explicitly for $s(x) \ne \bar{s}(x)$ distributions. 
Although the analysis of the above experiment seems to indicate 
small differences between the two distributions, the error bars are 
still large 
to be conclusive (see also discussion in Refs.~\cite{bro-ma,signal}).  

Here we present a novel procedure suitable to calculate 
quark and anti-quark non-perturbative distributions in the nucleon's sea. 
Within the framework of the meson cloud picture, we explicitly 
compute the probability density of having a meson and a baryon inside the 
nucleon, using splitting functions and recombination models.
The quark and anti-quark non-perturbative distributions in the parent
nucleon are obtained by means of the convolution of the meson and baryon
probabilities with the corresponding $s$ and $\bar{s}$ 
distributions inside baryon and meson respectively.

\section{The $q-\bar{q}$ asymmetry in the nucleon's sea}

We start by considering a simple picture of the nucleon in the 
infinite momentum frame as being formed by three dressed valence 
quarks so-called  valons\cite{hwa}, 
$
v(x) = \frac{105}{16} \sqrt{x} \left( 1 - x\right)^2.
$
In the framework of the meson cloud model, the nucleon can fluctuate to a  
meson-baryon bound state carrying zero {\it net} extra flavors. 
As a first step
in such a process, we may consider that each valon can emit a gluon which 
before interacting decays perturbatively into a $q\bar q$ pair. 
The probability of having such a perturbative $q\bar{q}$ pair can be 
evaluated in terms of the Altarelli-Parisi splitting functions, 
$ P_{gq} (z) = \frac{4}{3} \frac{1+(1-z)^2}{z},\ 
P_{qg} (z) = \frac{1}{2} \left( z^2 + (1-z)^2 \right),
\label{eq2}$
giving the probability 
of gluon emision and $q\bar{q}$ creation with momentum fraction $z$ 
from a parent quark and gluon respectively. Hence, 
\begin{equation}
q(x,Q^2) = \bar{q}(x,Q^2) = \frac{\alpha_{st}^2(Q^2)}{2\pi^2}
\int_{x+a}^1 {\frac{dy}{y} P_{qg}\left(\frac{x}{y}\right) 
\int_y^1{\frac{dz}{z} P_{gq}\left(\frac{y}{z}\right) v(z)}}
\label{eq3}
\end{equation}
is the joint probability density of obtaining a perturbative
quark or anti-quark 
coming from  subsequent decays $v \rightarrow v + g$ 
and $g \rightarrow q + \bar{q}$ at some fixed low  $Q^2$ scale. 
We have introduced a cut-off, $a$, in order to suppress the divercence 
at $x=0$ in the gluon distribution given by the inner integral
(see also Ref.\cite{we}). We fixed
it  by requiring that the quark distributions given by eq.~(\ref{eq3}) 
lay below experimental data \cite{grv} at a common scale. 
The range of values of $Q^2$ at which the process of virtual pair 
creation occurs is dictated by the valon model of the nucleon. For 
definiteness, we here use \cite{hwa} $Q = 0.8$ GeV, for which 
$\alpha_{st}$ is still sufficiently small to allow 
the perturbative evaluation of the $q\bar{q}$ pair production.
Once a $q\bar{q}$ pair is produced, it can rearrange itself with the 
remaining valons so as to form a most energetically favored 
meson-baryon bound state. The evaluation of meson and baryon formation 
has to be made using effective techniques in order to deal with 
the non-perturbative QCD processes involved in the dressing of quarks 
into hadrons. Although in-nucleon meson and baryon are virtual states, 
one may assume that the mechanisms involved in their formation are similar 
to those at work in the production of real hadrons in hadronic collisions;  
then, we can evaluate the $N\rightarrow MB$ fluctuation by means of a 
well-known recombination model approach. To ensure zero net 
extra flavors in the nucleon, as for momentum conservation, we may 
assume that the in-nucleon meson and baryon distributions satisfy 
$P_M(x) = P_B(1-x)$.  Now, we can for instance proceed 
to compute the strange meson distribution $P_M(x)$ along the lines of 
Ref.~\cite{das-hwa} and then relate it to the hyperon probability as 
indicated above (see Fig.1a) 
\footnote{The overall normalization is given by the probability 
that the strange hadronic fluctuation occurs; this value is commonly 
supposed to be around $4-10 \%$ (see {\it e.g.} Refs.\cite{bro-ma}).}.
Now, the non-perturbative strange and anti-strange sea distributions 
can be computed by means of two-level convolution formulas \cite{signal}
\be
s^{NP}(x) = \int^1_x {\frac{dy}{y} P_B(y)\ s_{B}(x/y)}\ \
\ \ \ \bar{s}^{NP}(x) = \int^1_x {\frac{dy}{y} P_M(y)\ \bar{s}_{M}(x/y)},
\label{eq8}
\ee
where the sources $s_{B}$ and $\bar{s}_{M}$ are primarily the 
momentum distributions of the valence quark and anti-quark 
in  baryon and meson respectively at the hadronic scale $Q^2$.
The most likely meson-baryon configurations are those closest to the 
nucleon energy shell, namely $\Lambda K$ and $\Sigma K$.

As long as experimental measurements are lacking at present,
several choices are possible for the $\bar{s}_M$ and $s_B$ 
distributions in $K$-ons and $\Lambda$, $\Sigma$ baryons respectively. 
In this respect, it is a common practice to employ modified light valence 
quark distributions of pions and protons \cite{signal}. However, 
we think that one should better use simple forms reflecting
the fact that strange quarks carry a rather large amount of momentum 
in $S=\pm 1$ hadron states at low $Q^2$. Indeed,
$\bar{s}_M(x)=6 x (1-x)$ adequately reflects this feature and, 
additionally, it reproduces quite well the momentum distribution found 
by Shigetani {\it et al.} \cite{shigetani} in the framework of a 
Nambu-Jona Lasinio model at low $Q^2$. For similar reasons, in $\Lambda$ 
and $\Sigma$ baryons we expect a $s$ momentum distribution peaked around 
$1/2$, which after normalization can be written as
$s_B(x)=12 x (1-x)^2$. The non-perturbative $s$ and $\bar{s}$ 
distributions 
of eqs.~\ref{eq8} as well as the corresponding $s-\bar{s}$ asymmetry are 
shown in Fig.~1b and 1c respectively.


\section{Summary and discussion}

We based our approach on a valon description of the ground state 
of the nucleon, which perturbatively produces sea quark/anti-quark pairs 
giving rise to a hadronic bound state by means of recombination with the 
remaining valons. In this way, a specific connection between the physics 
of hadronic reactions and that of hadron fluctuations is established. 
This is an important point of the approach since the physical 
principles related to confinement in QCD should be common to both processes.
It is interesting to note that neither  nucleon-meson-hyperon coupling 
constants nor vertex functions are needed for taking into account the 
extended nature of the nucleon-meson-hyperon vertex, avoiding the use of 
these controversial ingredients.

As we have shown in Fig.~1c, the model predicts a definite 
structured asymmetry in the strange sea of the nucleon. 
The shape of the $s$ and $\bar{s}$ distributions as well as 
their difference are similar to those found in a recent analysis  
based on a different approach \cite{bro-ma}. One should 
notice that the predicted non-perturbative quark and anti-quark 
distributions  depend on the form of the 
valence quark and anti-quark distributions of the in-nucleon 
baryon and meson respectively. So, although the model is reliable in 
predicting different distributions for sea quark and anti-quarks, their 
exact shapes remain unknown until we have more confident results for  
valence densities inside strange hadrons. Nevertheless, the model predicts 
an excess of quarks over anti-quarks in the nucleon sea carrying a 
large fraction of the nucleon's momentum (see Fig.~1c) and
this prediction appears to be independent of the exact form of the strange 
and anti-strange distributions inside the in-nucleon baryon and meson. 
\begin{figure}[t]
\begin{minipage}{2.5in}
\psfig{figure=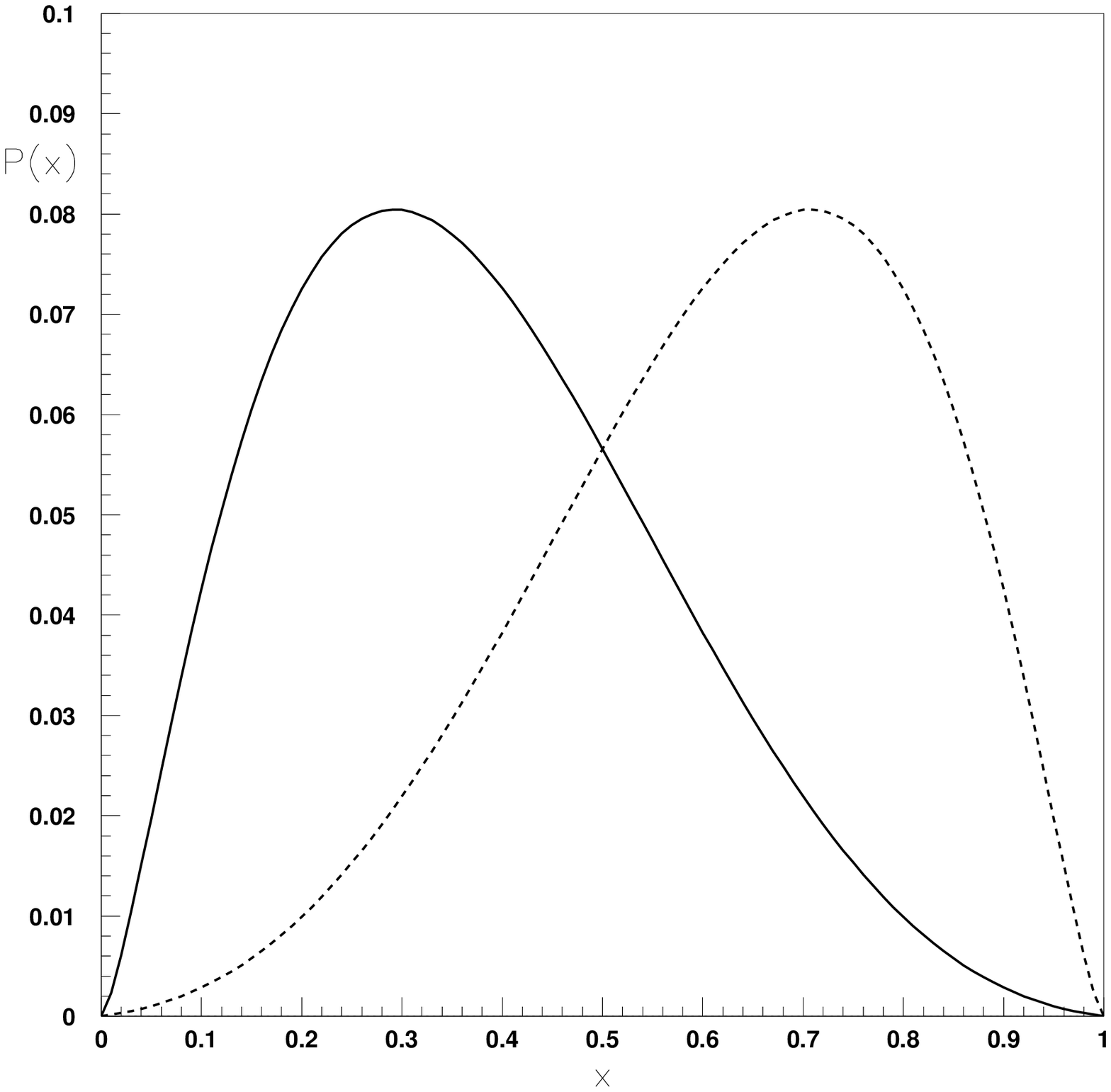,height=2.3in}
\end{minipage}
\begin{minipage}{7.5in}
\psfig{figure=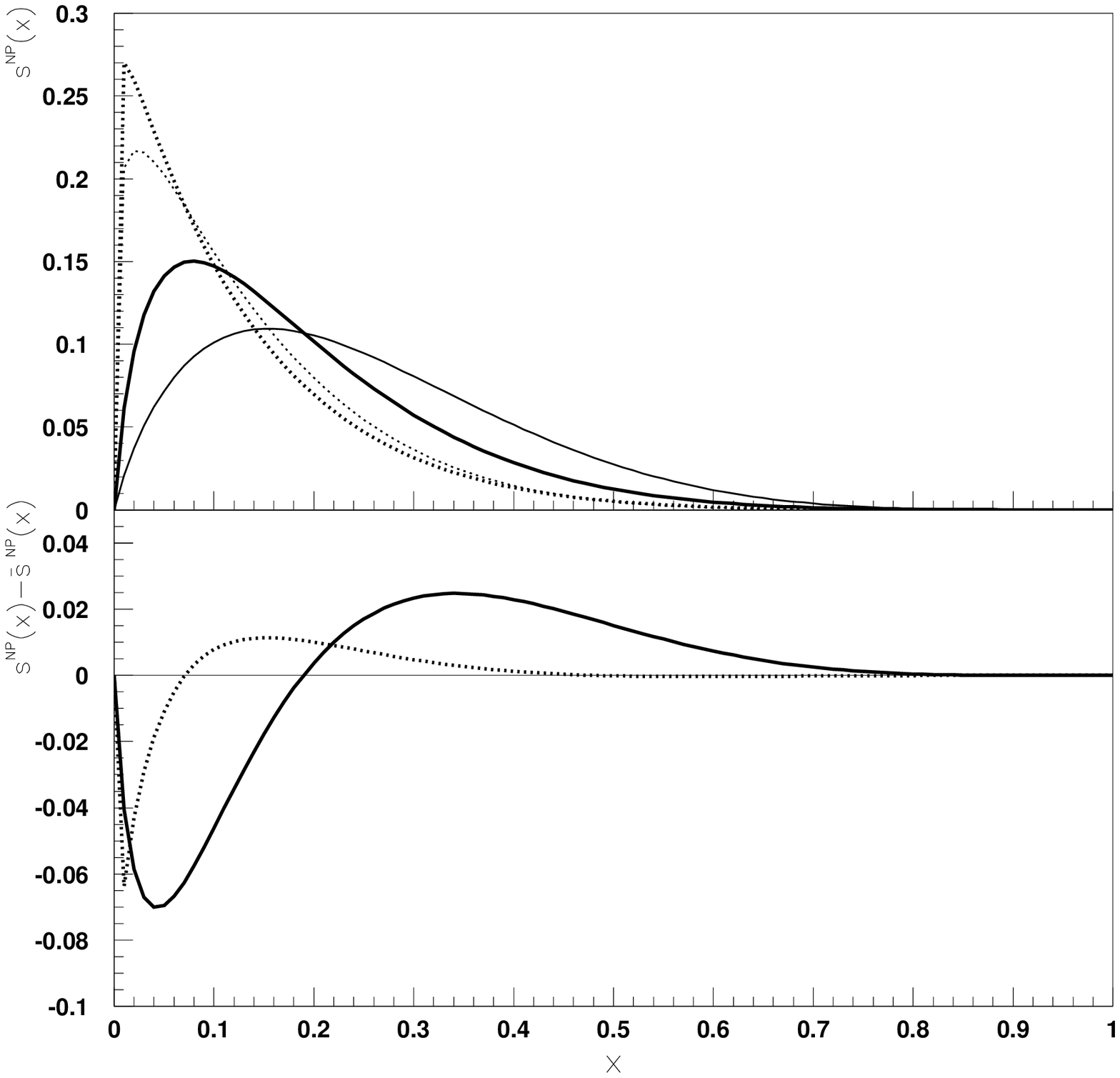,height=2.3in}
\end{minipage}
\caption{Left: a) Meson (full line) and baryon (dashed line) probability 
densities in the nucleon. Right: b) 
Full lines are $s^{NP}$ (thin) and $\bar{s}^{NP}$ 
(thick), as obtained by using our simple forms for $s_B$ and 
$\bar{s}_M$. 
Point lines result of using, instead,  light-like distributions.
c) The corresponding $s(x)-\bar{s}(x)$  asymmetries. See Ref.~5 for more 
details.
\label{fig:radish}}
\end{figure}

\section*{Acknowledgments} 

The authors are supported by FAPERJ, Rio de Janeiro. 
H.R.C. would like to acknowledge the 
Organizing Comitee of {\it QM'97} and {\it Hadrons'98} for warm hospitality
and financial support. 


\clearpage

\section*{References}
\begin{thebibliography}{99}

\bibitem{ccfr} A.O. Bazarko {\it et al.} (CCFR Collaboration), 
\Journal{\ZPC} {65}{189}{1995}.

\bibitem{bro-ma} S.J. Brodsky and B.Q. Ma, \Journal{\PLB}{381} 
{317} {1996}. 

\bibitem{signal} A. Signal and A.W. Thomas, 
\Journal{\PLB} {191}{205}{1987}.

\bibitem{hwa} R.C. Hwa, \Journal{\PRD} {22}{759}{1980}; 
{\it ibid.} 1593.

\bibitem{we} H.R. Christiansen and J. Magnin, {\it Strange 
anti-strange asymmetry in the nucleon sea}, hep-ph/9801283,  
to appear in {\it Phys. Lett}~ B.


\bibitem{grv} M. Gl\"uck, E. Reya and A. Vogt, \Journal{\ZPC} 
{53}{127}{1992}.

\bibitem{das-hwa} K.P. Das and R.C. Hwa, \Journal{\PLB} {68}{459}{1977}.
\nopagebreak
\bibitem{shigetani} T. Shigetani, K. Suzuki and H. Toki, 
\Journal{\PLB} {308}{383}{1993}. 
J.T. Londergan {\it et al.} \Journal{\PLB} {380}{393}{1996}.

\end{thebibliography}
\end{document}